\documentclass[fleqn,usenatbib]{mnras}

\usepackage{newtxtext,newtxmath}

\usepackage[T1]{fontenc}

\DeclareRobustCommand{\VAN}[3]{#2}
\let\VANthebibliography\thebibliography
\def\thebibliography{\DeclareRobustCommand{\VAN}[3]{##3}\VANthebibliography}

\usepackage{graphicx}	
\usepackage{amsmath}	

\title[All along the line of sight]{All along the line of sight: a closer look at opening angles and absorption regions in the atmospheres of transiting exoplanets}

\author[Joost P. Wardenier et al.]{
Joost P. Wardenier,$^{1}$\thanks{E-mail: joost.wardenier@physics.ox.ac.uk}
Vivien Parmentier,$^{1}$
Elspeth K.H. Lee$^{2}$
\\
$^{1}$Department of Physics (Atmospheric, Oceanic and Planetary Physics), University of Oxford, Oxford, OX1 3PU, UK\\
$^{2}$Center for Space and Habitability, University of Bern, Gesellschaftsstrasse 6, CH-3012 Bern, Switzerland
}

\date{Accepted XXX. Received YYY; in original form ZZZ}

\pubyear{2015}

\begin{document}
\label{firstpage}
\pagerange{\pageref{firstpage}--\pageref{lastpage}}
\maketitle

\begin{abstract}
Transmission spectra contain a wealth of information about the atmospheres of transiting exoplanets. However, large thermal and chemical gradients along the line of sight can lead to biased inferences in atmospheric retrievals. In order to determine how far from the limb plane the atmosphere still impacts the transmission spectrum, we derive a new formula to estimate the opening angle of a planet. This is the angle subtended by the atmospheric region that contributes to the observation along the line of sight, as seen from the planet centre. We benchmark our formula with a 3D Monte-Carlo radiative transfer code and we define an opening angle suitable for the interpretation of JWST observations, assuming a 10-ppm noise floor. We find that the opening angle is only a few degrees for planets cooler than ca. 500 Kelvins, while it can be as large as 25 degrees for (ultra-)hot Jupiters and 50 degrees for hot Neptunes. Compared to previous works, our more robust approach leads to smaller estimates for the opening angle across a wide range scale heights and planetary radii. Finally, we show that ultra-hot Jupiters have an opening angle that is smaller than the angle over which the planet rotates during the transit. This allows for time-resolved transmission spectroscopy observations that probe independent parts of the planetary limb during the first and second half of the transit.
\end{abstract}

\begin{keywords}
radiative transfer -- methods: numerical -- planets and satellites: atmospheres -- planets and satellites: gaseous planets
\end{keywords}



\section{Introduction}
\label{sect:intro}

Over the past two decades, transmission spectroscopy (e.g., \citealt{Seager2000, Charbonneau2002, Snellen2008, Sing2016, Nikolov2018}) has been integral to the atmospheric characterisation of transiting exoplanets -- constraining their chemical abundances and temperature profiles. When an exoplanet passes in front of its host star, a small fraction of the stellar photons are absorbed and scattered by the planetary atmosphere. Because the atmosphere's opacity is wavelength-dependent, measuring the effective radius of the planet $R_\text{eff}(\lambda)$ at different wavelengths results in a transmission spectrum. The transmission spectrum is generally expressed in terms of a depth $\delta(\lambda) = R_\text{eff}(\lambda)^2/R_{\star}^2$, with $R_{\star}$ the radius of the host star.  

In order to physically interpret a transmission spectrum and infer the properties of the underlying atmosphere, it is essential to know what regions of the atmosphere are actually probed by the observation. In 1D radiative-transfer models, it is possible to compute a so-called \emph{transmission contribution function} (TCF, \citealt{Barstow2013,Molliere2019a}), which quantifies how substantially an atmospheric layer contributes to the spectrum at a given wavelength. The basic idea is to switch off (or perturb) the opacities layer-by-layer and compute the resulting transit radius of the planet. The greater the deviation from the original radius, the greater the contribution by the associated layer. In this way, the TCF can be seen as the derivative of the transmission with respect to pressure.

Because exoplanet atmospheres are 3D, the atmospheric region probed by transmission spectroscopy $-$ henceforth called the \mbox{\emph{absorption region}} $-$ does not only extend along the altitude axis, but also along the line of sight, perpendicular to the limb plane. 1D models that are typically used to interpret observations ignore this dimension (e.g., \citealt{Line2013,Waldmann2014,Molliere2019a}). However, a collection of recent works (\citealt{Caldas, Lacy2020, Pluriel2020, Pluriel2021}) have shown that thermal and chemical gradients between the dayside and nightside of tidally locked gas giants can have a significant impact on their transmission spectrum -- for earlier studies that computed spectra from 3D models, see \citet{Burrows2010} and \citet{Fortney2010}. Therefore, it is valuable to assess \emph{how far} from the limb plane the atmosphere still influences a planet's transmission spectrum, and thus \emph{how stretched} the absorption region is along the line of sight. Essentially, we require a TCF for the line-of-sight dimension.

The absorption region can be seen as a ``resolution element''. If the absorption region is compact, the transmission spectrum probes a very specific part of the atmosphere, with its own temperature and chemistry. On the other hand, if the absorption region is stretched along the line of sight, the transmission spectrum is shaped by the atmospheric structure across many different longitudes. In the context of tidally locked planets, this has important implications. Ultra-hot Jupiters (\citealt{Arcangeli2018, Bell2018, Parmentier2018}), for example, rotate by tens of degrees during a transit, owing to their short orbital radii.\footnote{Under the assumption of tidal locking, WASP-76b and WASP-121b (two typical ultra-hot Jupiters) rotate by 31 and 35 degrees during their transit, respectively. This includes the ingress and egress phase.} Hence, one can expect observations to probe different longitudes at different orbital phases (e.g., \citealt{Ehrenreich2020, Hoeijmakers2020, Bourrier2020, Borsa2021, Kesseli2021,Wardenier2021}). However, it is the resolution element that determines how independent the measurements (of wind speeds, temperature and composition) at the various orbital phases are. The size of the absorption region governs how much local information can be retrieved from the transmission spectrum. 

The goal of this work is to obtain an estimate for the extent of the absorption region along the line of sight -- and thus assess the importance of accounting for this dimension when interpreting observations. In Section \ref{sect:prev_work}, we briefly discuss work that has already been done in this regard. In Section \ref{sect:derivation}, we derive a new analytical formula for the opening angle (i.e. the angular extent of the absorption region along the line of sight) and we validate it using a Monte-Carlo radiative transfer code. In Section \ref{sect:wavelength_dependence}, we study the wavelength dependence of the absorption region and the opening angle. In Section \ref{sect:noise_floor}, we assess how the opening angle behaves when an instrument noise floor is assumed. Finally, we discuss the implications of this work in Section \ref{sect:discussion}, followed by a conclusion in Section \ref{sect:conclusion}.

\section{Previous Work}
\label{sect:prev_work}

\begin{figure}
\centering
\includegraphics[width=0.47\textwidth]{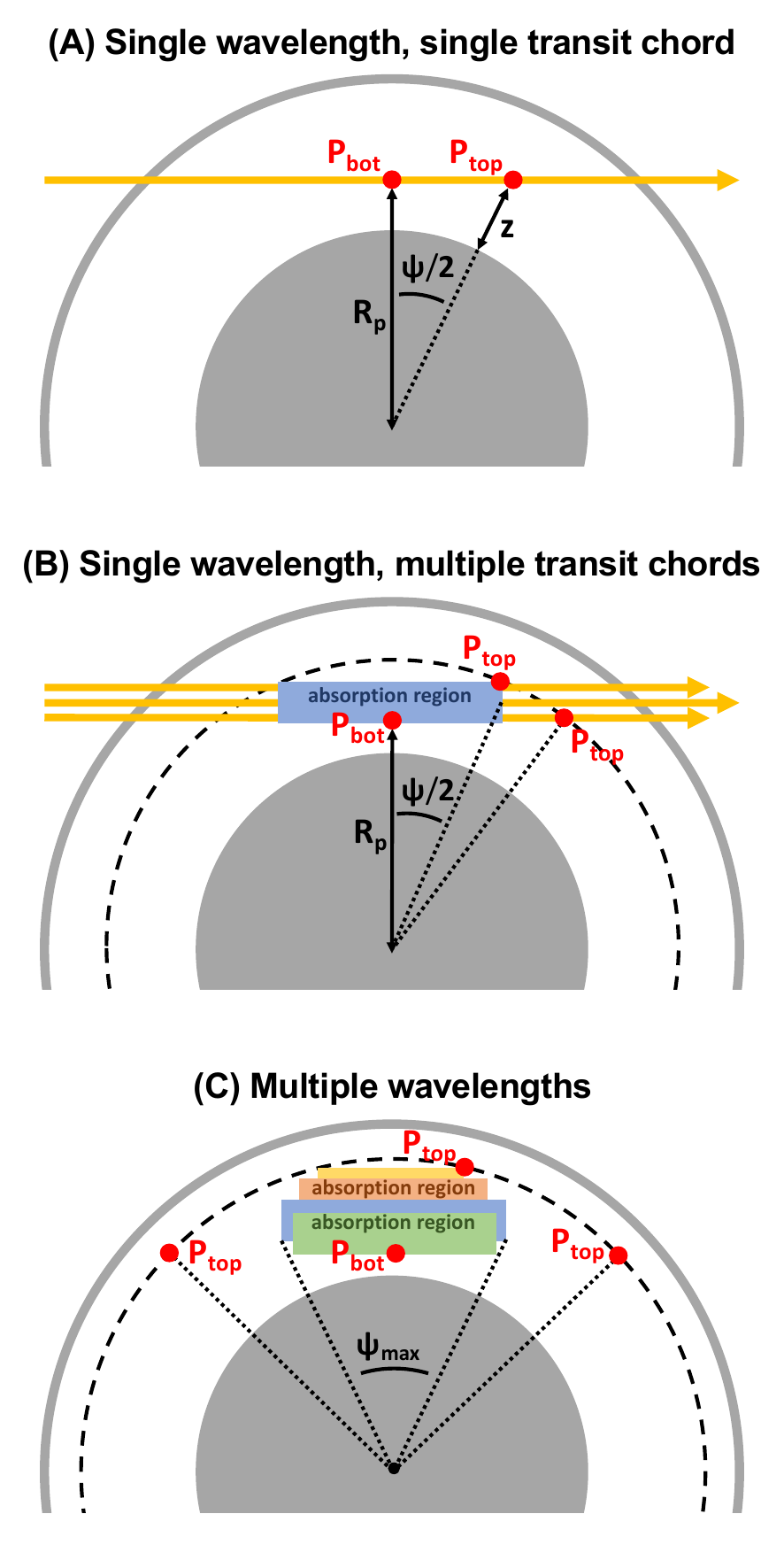}
\vspace{-13pt}
\caption{Cross section of a transiting exoplanet, with the host star on the left and the observer on the right. The grey disk represents the planetary interior, while the white annulus is the atmosphere of the planet. \textbf{(A)} For a stellar light ray with a particular impact parameter $R+z_t$ (yellow arrow), the opening angle $\psi$ can be computed from trigonometry, using the maximum pressure $P_\text{bot}$ and minimum pressure $P_\text{top}$ probed by the light ray. For pressures lower than $P_\text{bot}$, the atmosphere's impact on the transmission spectrum is negligible. \textbf{(B)} In reality, the absorption region probed by the observation at a particular wavelength is vertically extended, such that $P_\text{bot}$ and $P_\text{top}$ do not lie along the same transit chord. \textbf{(C)} The absorption regions associated with different wavelengths lie at different altitudes. The values of the highest $P_\text{bot}$ and the lowest $P_\text{top}$ can suggest a much larger opening angle than is truly the case.}
\label{fig:opening_geometry}
\end{figure}

\citet{Caldas} used a geometrical argument to estimate what portion of the atmosphere is probed along the line of sight. To this end, they introduced the opening angle $\psi$. This is the angle (as seen from the centre of the planet) subtended by the region that contributes to the observation. The top panel of Fig. \ref{fig:opening_geometry} illustrates how $\psi$ can be computed. Firstly, one has to assume a minimum pressure $P_\text{top}$ and a maximum pressure $P_\text{bot}$ between which the atmospheric opacities have a ``measurable'' impact on the planet's transmission spectrum (analogous to the TCF). The opening angle can then be defined as 

\begin{equation}
    \psi \equiv 2 \arccos \bigg( \frac{R_\text{p}}{R_\text{p} + z(P_\text{top})} \bigg),
\end{equation}

\noindent with $R_\text{p}$ the planetary radius at $P_\text{bot}$ and $z(P_\text{top})$ the altitude at $P_\text{top}$. Assuming an isothermal atmosphere with variable gravity, the hydrostatic equation reads

\begin{equation}
    \int_{P_\text{bot}}^{P_\text{top}} \frac{dP}{P} = - \int_{0}^{z(P_\text{top})} \frac{dz}{H} \bigg( \frac{R_\text{p}}{R_\text{p} + z} \bigg)^2,
\end{equation}

\noindent where $H$ is the scale height at $R_\text{p}$. Solving this equation yields

\begin{equation}
    z(P_\text{top}) = \frac{R_\text{p}}{1-\frac{H}{R_\text{p}}\ln \Big(\frac{P_\text{bot}}{P_\text{top}}\Big)} - R_\text{p},
\end{equation}

\noindent such that the opening angle becomes (\citealt{Caldas})

\begin{equation} \label{eq:def_opening_angle}
    \psi \equiv 2 \arccos \bigg( 1 -  \frac{H}{R_\text{p}}\ln \bigg(\frac{P_\text{bot}}{P_\text{top}}\bigg) \bigg).
\end{equation}

\noindent This result demonstrates that the opening angle is mainly governed by the ratio between the scale height and the radius of the planet. The greater this ratio, the larger the opening angle and the greater the importance of the line-of-sight dimension. Based on arguments from the literature, \citet{Caldas} assumed $(P_\text{bot}, P_\text{top}) = (10^{-2}, 10^{-5})$ bar. They found that the opening angle of the hot Jupiter \mbox{HD 209458 b} can be expected to exceed 30 degrees, while that of the sub-Neptune \mbox{GJ 1214 b} lies in the order of 45--50 degrees.



Although equation \ref{eq:def_opening_angle} provides a quick and simple estimate for the opening angle, one downside is the assumption that the minimum and maximum pressure probed by the observation are situated along the same transit chord (top panel in Fig. \ref{fig:opening_geometry}). In this case, the triangle with $P_\text{top}$ and $P_\text{bot}$ as two of its vertices is right-angled. However, if we consider a particular wavelength, the absorption region does not only stretch along the line of sight, but also in the vertical direction. As shown in the middle panel of Fig. \ref{fig:opening_geometry}, the minimum and maximum pressure contained in this absorption region do not lie along the same transit chord. As a result, the opening angle that is found by plugging $P_\text{bot}$ and $P_\text{top}$ into equation \ref{eq:def_opening_angle} is larger than the actual angular extent of the absorption region. This effect is enhanced when we consider multiple wavelengths. In a transmission spectrum, the effective radius of the planet changes because the absorption region moves up and down with wavelength. Consequently, the overall highest $P_\text{bot}$ and the overall lowest $P_\text{top}$ probed by the observation lie even further apart compared to the single-wavelength case (because the pressures are not probed by the same wavelength). This is illustrated in the bottom panel of Fig. \ref{fig:opening_geometry}. Because $P_\text{top}$ and $P_\text{bot}$ lie on two different transit chords, the opening angle is overestimated.

\section{A new formula for the opening angle}
\label{sect:derivation}

In this section, we revisit the opening angle introduced by \citet{Caldas} and we derive an alternative formula that does not contain $P_\text{top}$ and $P_\text{bot}$. Up to equation \ref{eq:z_angle}, our approach is similar to the derivation presented in Section A.2 from \citet{Caldas}. However, the authors do not explicitly solve for the opening angle of a uniform, 1D atmosphere, as we do in equation \ref{eq:opening_angle_final_result}. Additionally, we use a radiative-transfer code to validate the new formula and explicitly reconstruct the absorption regions discussed in the previous sections.

\subsection{Derivation}

We consider an isothermal atmosphere with one absorbing species and a transit chord at a height $z_\text{t}$ above the reference radius $R_\text{p}$. Under the assumption of a uniform composition and a constant gravity, the integrated optical depth $\tau$ at position $x$ is given by (\citealt{Caldas}, Appendix A): 

\begin{equation} \label{eq:tau_caldas}
    \tau(x, z_\text{t}) = \sqrt{2\pi R_\text{p} H} \ \sigma_\text{mol} \ \chi \ n_0 \ e^{-z_\text{t}/H} \bigg[ \frac{1}{2} + \frac{1}{2} \text{erf} \bigg( {\frac{x}{\sqrt{2 R_\text{p} H}}} \bigg) \bigg],
\end{equation}

\noindent with $\tau$ and $x$ increasing in the direction towards the observer and $x = 0$ in the limb plane. Additionally, $\sigma_\text{mol}$ is the constant\footnote{In reality, $\sigma_\text{mol}$ is a function of pressure, temperature and wavelength.} cross-section of the considered species, $\chi$ is its number fraction, $n_0$ is the particle number density at the reference radius, and $H$ is the atmosphere's constant scale height. The derivation of equation \ref{eq:tau_caldas} also requires that $|z_\text{t}| \ll 2R_\text{p}$, such that the altitude at position $x$ can be written as $z \approx z_\text{t} + x/2R_\text{p}$ (\citealt{Fortney2005, Caldas}). In other words, the distance between the transit chord and the reference radius should be much smaller than the reference radius itself.

\begin{figure}
\centering
\vspace{-11pt}
\includegraphics[width=0.5\textwidth]{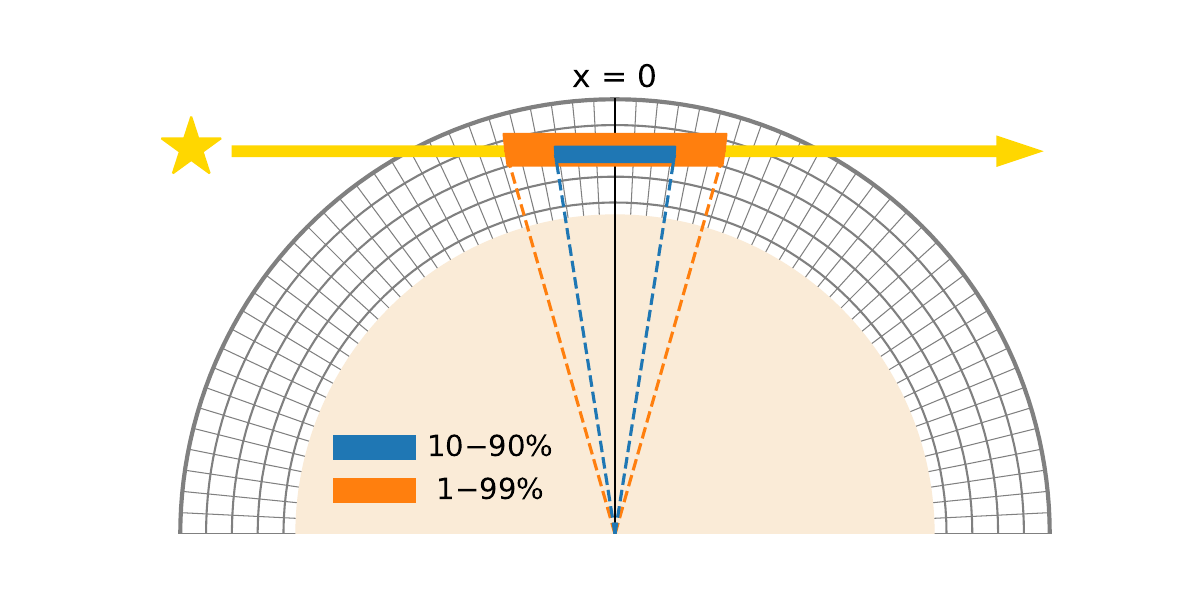}
\vspace{-13pt}
\caption{10$-$90\% and 1$-$99\% absorption regions for the grey atmosphere with $R_\text{p}$ = 10 $R_\text{Earth}$ and $H$ = 1000 km, computed with \textsc{hires-mcrt}. The limb of the planet is indicated by $x = 0$, while the yellow arrow represents a stellar light ray crossing the atmosphere. The grey circles represent isobars of 1, 10$^{-2}$, 10$^{-4}$, 10$^{-6}$ and 10$^{-8}$ bar (in reality, the models have 100 pressure layers). The dashed lines show the opening angles associated with both absorption regions. The atmosphere and the planet are to scale in this figure.}
\label{fig:absorption_region}
\end{figure}

When we integrate along the full transit chord, the total optical depth becomes

\begin{equation}
    \tau_0(z_\text{t}) \equiv \tau(x \rightarrow \infty, z_\text{t}) = \sqrt{2\pi R_\text{p} H} \ \sigma_\text{mol} \ \chi \ n_0 \ e^{-z_\text{t}/H}.
\end{equation}

\noindent This equation allows us to express the altitude associated with the transit chord as a function of its total optical depth $\tau_0$ (see also \citealt{Etangs2008}):

\begin{equation} \label{eq:z_angle}
    z_\text{t}(\tau_0) = H \ln \bigg( \sqrt{2\pi R_\text{p} H} \ \sigma_\text{mol} \ \chi \ n_0 /\tau_0  \bigg).
\end{equation}

\noindent We now define $\beta$, the ratio between the optical depth at a certain position $x$ and the total optical depth associated with the transit chord:

\begin{equation}
    \beta(x) \equiv \frac{\tau(x, z_\text{t})}{\tau_0(z_\text{t})} = \frac{1}{2} + \frac{1}{2} \text{erf} \bigg( {\frac{x}{\sqrt{2 R_\text{p} H}}} \bigg),
\end{equation}

\noindent which is independent of $z_t$. Inverting this equation yields

\begin{equation} \label{eq:x_angle}
    x(\beta) = \sqrt{2 R_\text{p} H} \ \text{erf}^{-1}(2\beta-1).
\end{equation}

\begin{figure*}
\vspace{-10pt}
\centering
\makebox[\textwidth][c]{\includegraphics[width=0.9\textwidth]{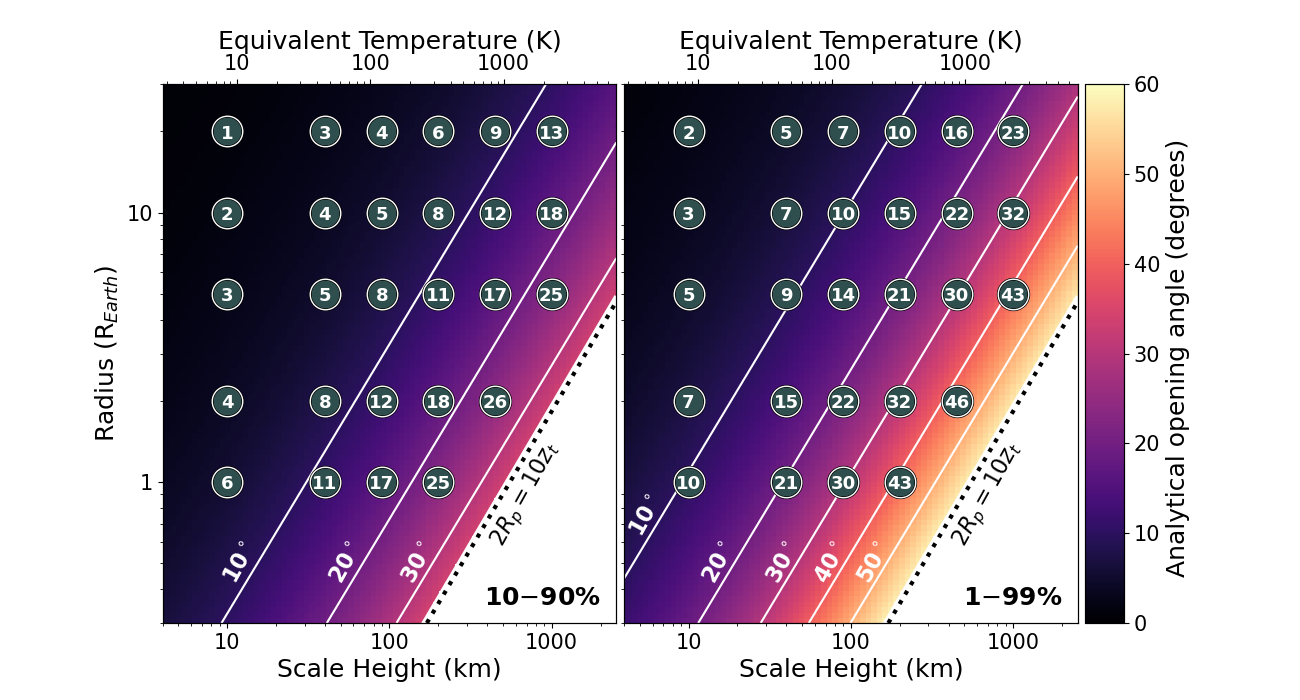}}
\vspace{-15pt}
\caption{Colour maps showing the opening angle from equation \ref{eq:opening_angle_final_result} as a function of scale height $H$ and planetary radius $R_\text{p}$, for $\beta = 0.1$ (left panel, 10$-$90\% absorption) and $\beta = 0.01$ (right panel, 1$-$99\% absorption). Additionally, we used $\chi \sigma_\text{mol}/\tau_0$ = $10^{-24.3}$ cm$^{-2}$. In the region to the right of the dotted lines, the approximation $|z_\text{t}| \ll 2R_\text{p}$ starts to break down, and hence it was excluded from the map. The values in the grey circles denote the opening angles (in degrees) that were geometrically calculated from the Monte-Carlo radiative transfer.}
\label{fig:formula_verification}
\end{figure*}

\noindent Using equations \ref{eq:z_angle} and \ref{eq:x_angle}, we can now compute the opening angle associated with a transit-chord segment along which the optical depth increases from a fraction $\beta$ to a fraction $(1 - \beta)$ of the total optical depth $\tau_0$:

\begin{align}
\begin{split} \label{eq:opening_angle_final_result} 
    \psi(\tau_0, \beta) = {} & 2 \arctan \Bigg( \frac{|x(\beta)|}{ R_\text{p} + z_\text{t}(\tau_0)} \Bigg) \\
    = {} & 2 \arctan \Bigg( \frac{\sqrt{2R_\text{p}H} \ \text{erf}^{-1}(1-2\beta)}{ R_\text{p} + H \ln \big( \sqrt{2\pi R_\text{p} H} \ \sigma_\text{mol} \ \chi \ n_0 /\tau_0  \big) } \Bigg),
\end{split}
\end{align}

\noindent with $\beta < 0.5$. The argument of the inverse error function was negated to get rid of the absolute-value symbols. To compute the opening angle subtended by an absorption region (i.e. the region of the atmosphere about which the transmission spectrum contains information), one can set $\tau_0 =$ 0.56 (\citealt{Etangs2008}). For a wide range of $R_\text{p}/H$ values, a transit chord with this optical depth lies at an altitude $z_\text{t}$ such that a fully opaque disk with radius $R_\text{p} + z_\text{t}$ would give rise to the same transit depth as the planet. About 43\% of the light is absorbed along a transit chord with $\tau_0 =$ 0.56.

\subsection{Numerical verification}
\label{sect:num_ver_formula}

\subsubsection{Atmospheric models and radiative transfer}
\label{sect:radtrans1}

To verify whether equation \ref{eq:opening_angle_final_result} produces the correct values for $\psi$ (given all the simplifying assumptions), we set up 1D isothermal atmospheres for 27 combinations of scale height \mbox{$H \in \{10, 40, 90, 200, 450, 1000\}$ km} and planetary radius \mbox{$R_\text{p} \in \{1, 2, 5, 10, 20\}$ $R_\text{Earth}$}. For each $(H, R_\text{p})$ combination, we assume a constant gravity $g = 10$ m/s$^2$ and mean molecular weight \mbox{$\mu$ = 2.3 $m_\text{H}$}. In accordance with \citet{Caldas}, we define the planetary radius to coincide with $P = 0.01$ bar. The atmospheres have 100 vertical layers, over which the pressure drops from $P = 10$ bar (bottom) to $P = 10^{-8}$ bar (top). Additionally, the (equivalent) temperature of the atmospheres follows from $T = \mu H g / k_B$, with $k_B$ the Boltzmann constant. The particle number density at the reference radius, $n_0$, follows from the ideal gas law. 

In each model, we assume that there is one absorber with a constant abundance and cross-section, such that $\chi \sigma_\text{mol}/\tau_0 = 10^{-24.3}$ cm$^{-2}$ (corresponding to e.g., $\chi = 10^{-3.3}$,  $\sigma_\text{mol} = 10^{-21.3}$ cm$^{-2}$ and $\tau_0 = 0.56$). These values are typical for an atmosphere with H$_\text{2}$O and/or CO at a solar abundance level (e.g., \citealt{Venot2015,Line2021}), but they could represent any chemical species -- whichever one turns out to be the dominant absorber. In Appendix \ref{ap:A}, we show that the dependence of the opening angle on $\chi \sigma_\text{mol}/\tau_0$ is relatively weak (because of the logarithm in equation \ref{eq:opening_angle_final_result}), so a rough estimate suffices. On the other hand, changing the value of $\beta$ has a much bigger effect on the opening angle.

Once all atmospheric models have been obtained, we compute their absorption regions. To this end, we use \textsc{hires-cmcrt} (\citealt{Wardenier2021}), a high-resolution version of the cloudy Monte-Carlo radiative transfer code (\textsc{cmcrt}) developed by \citet{Lee2017, Lee2019}. Since the opacity of the atmospheres is constant, their transmission spectrum is flat and it suffices to compute the absorption region at a single wavelength point. For the purposes of this work, we use \textsc{hires-mcrt} as a 2D code, where we only illuminate the equatorial plane of the planet. We initialise $10^5$ photon packets with a random impact parameter $R_\text{p} + z_\text{t} \in$ \mbox{$(R_\text{p} + z_\text{min}, R_\text{p} + z_\text{max})$}, where $z_\text{min} < 0$ and $z_\text{max} > 0$ are the altitudes at the bottom and the top of the atmosphere, respectively. For each photon packet, we evaluate the optical depth along its transit chord in a polar geometry, with 100 vertical levels and 128 longitude slices. That is,

\begin{equation} \label{eq:tau_grey}
\tau_0(z_t) = \chi \sigma_\text{mol} \sum_{i=1}^{N_{\text{cells}}} n_i \Delta x_i,
\end{equation}

\noindent with $n_i$ the particle number density in the $i$-th atmospheric cell and $\Delta x_i$ the distance the photon packet travels through this cell.




\subsubsection{Constructing absorption regions}
\label{sect:absorption-regions}

In addition to equation \ref{eq:tau_grey}, we also keep track of the optical depth encountered by a photon packet as a function of position. Using this information, we can reconstruct the absorption regions of the atmosphere. For a given value of $\beta$ (see equation \ref{eq:opening_angle_final_result}), we define the absorption region in two steps. Firstly, we select all transit chords for which the total transmission $e^{-\tau_0}$ ranges between $\beta$ and \mbox{$1-\beta$}. Secondly, for each transit chord from the first step, we select the segment along which the optical depth increases from \mbox{$\beta\tau_0$} to \mbox{$(1-\beta)\tau_0$}. All segments together constitute the absorption region.\footnote{For instance, when $\beta = 0.1$, the absorption region includes all transit chords for which $0.1 < e^{-\tau_0} < 0.9$. Additionally, we only select the segment of every transit chord where the optical depth increases from 10\% to 90\% of its final value, such that the absorption region does not stretch all the way to the (arbitrary) model boundaries. By construction, the absorption region is symmetric about the limb plane in the limit of a 1D, uniform atmosphere.} We compute absorption regions for $\beta = 0.01$ and $\beta = 0.1$, which will be referred to as the 1$-$99\% and 10$-$90\% absorption regions in the rest of this work.

Fig. \ref{fig:absorption_region} shows the absorption regions for the model with \mbox{$R_\text{p}$ = 10 $R_\text{Earth}$} and \mbox{$H$ = 1000 km}. The 10$-$90\% region is fully contained by the 1$-$99\% region, in agreement with the definition. Once the absorption regions are computed, their corresponding opening angle $\psi_{\text{abs}}$ can be found geometrically (see Fig. \ref{fig:absorption_region}):

\begin{equation} \label{fig:angle_abs_region}
    \psi_{\text{abs}} = 2 \arctan \bigg(\frac{\bar{x}}{2\bar{y}} \bigg),
\end{equation}

\noindent with $\bar{x}$ the average length of the segments in the absorption region and $\bar{y}$ the average distance of the segments from the centre of the planet. Equation \ref{fig:angle_abs_region} holds as long as the absorption region is symmetric about the limb plane.

\subsubsection{Comparing the formula to the model}

Fig. \ref{fig:formula_verification} shows the opening angles computed from equation \ref{fig:angle_abs_region} for each of the 27 atmospheric models. For both values of $\beta$, there is a good agreement between the analytical opening angle from equation \ref{eq:opening_angle_final_result} (colour map) and the opening angle inferred from the absorption regions computed with the Monte-Carlo code (values in grey circles). It should be noted that the \emph{analytical} opening angle is only based on the transit chord with $\tau_0 = 0.56$, while the opening angles associated with the absorption regions are based on all transit chords with $\beta < e^{-\tau_0} < 1 - \beta$. This may explain slight differences in Fig. \ref{fig:formula_verification} towards higher values of $\psi$. 

The trends in Fig. \ref{fig:formula_verification} are very similar to those in Fig. 2 from \citet{Caldas}. The value of the opening angle increases as a function of the ratio $H/R_\text{p}$, which is in agreement with equation \ref{eq:def_opening_angle}. Furthermore, the plots illustrate how the value of $\beta$ impacts the opening angle. For the largest scale heights, the 1--99\% opening angles are a factor $\sim$1.8 bigger than the 10--90\% opening angles. This is because with increasing $\beta$, a longer segment of the transit chord is included in the absorption region. In Section \ref{sect:noise_floor}, we will investigate which value of $\beta$ yields a realistic estimate for the true opening angle of a planet. 

\section{The wavelength dependence of the opening angle}
\label{sect:wavelength_dependence}

In reality, planet atmospheres do not have a constant opacity. In this section, we examine how the opening angle changes when we account for the fact that $\sigma_{\text{mol}}$ is a function of pressure, temperature and wavelength.  

\subsection{Atmospheric models and radiative transfer}
\label{sect:radtrans2}

\begin{figure}
\vspace{-45pt}
\centering
\includegraphics[width=0.5\textwidth]{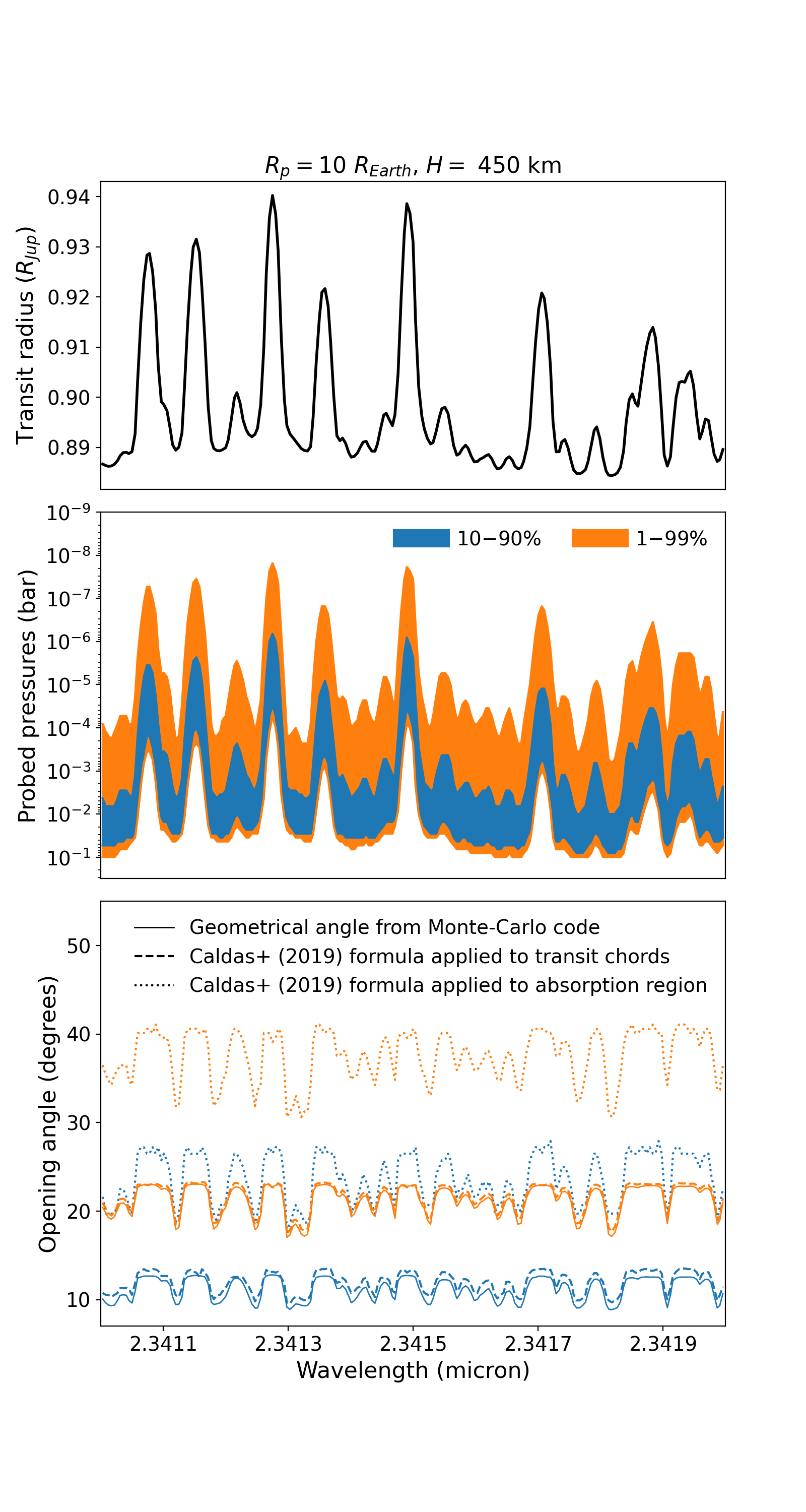}
\vspace{-52pt}
\caption{Radiative-transfer output for the model with $R_\text{p} = 10$  $R_\text{Earth}$ and $H = 450$ km. \textbf{Top:} The transmission spectrum, featuring H$_\text{2}$O and CO absorption lines. \textbf{Middle:} The pressures contained by the 10$-$90\% and 1$-$99\% absorption regions as a function of wavelength. \textbf{Bottom:} The 10$-$90\% and 1$-$99\% opening angles as a function of wavelength. The solid curves were obtained using the method described in section \ref{sect:absorption-regions}. The dashed curves show the output of the \citet{Caldas} formula when the $P_\text{bot}$ and $P_\text{top}$ of \emph{individual} transit chords are plugged into the formula, while the dotted curves illustrate what happens when the $P_\text{bot}$ and $P_\text{top}$ of the full absorption region are used (as depicted in the middle panel of Fig. \ref{fig:opening_geometry}). The latter leads to an overestimation of the opening angle.}
\label{fig:three_panels}
\end{figure}

We set up 27 atmospheric models with the same parameters as those described in Section \ref{sect:radtrans1}. This time, however, we assume a variable gravity (\mbox{$g =$ 10 m/s$^2$} at the reference radius), in accordance with the formula from \citet{Caldas}. Furthermore, we include the chemical species H$_\text{2}$, He, H$_\text{2}$O and CO, with number fractions $\chi_\text{H2} = 0.76$, $\chi_\text{He} = 0.24$, $\chi_\text{H2O} = 10^{-3.3}$ and $\chi_\text{CO} = 10^{-3.3}$. Again, we use \textsc{hires-mcrt} as a 2D code to compute the absorption regions and transit spectra associated with the models. We account for the opacities of H$_2$O (\citealt{Polyansky2018}) and CO (\citealt{Li2015}), taken from the \texttt{ExoMol} database (\citealt{Tennyson2016, Tennyson2020}), as well the continuum due H$_2$-H$_2$ and H$_2$-He collision-induced absorption (\citealt{Borysow2001, Borysow2002, Gordon2017}). The radiative transfer is performed at high spectral resolution (\mbox{$R = 500,000$}), for 213 wavelength points near 2.34 micron (see Fig. \ref{fig:three_panels}). The rationale behind this approach is that at high resolution, cross-sections vary dramatically as a function of wavelength $-$ depending on whether $\lambda$ lies inside or outside a line core. Therefore, modelling the atmospheric absorption around just a number of line cores should already produce a spectrum that probes many different atmospheric layers. 

Because the cross-sections are no longer constant, we now evaluate the optical depth along a transit chord as:

\begin{equation}
\tau_0(z_t, \lambda) = \sum_{i=1}^{N_{\text{cells}}} n_i \Delta x_i \Bigg[ \sum_{j=1}^{N_{\text{species}}} \chi_j \sigma_{\text{mol}, i, j}(\lambda) \Bigg],
\end{equation}

\noindent with $\chi_j$ the number fraction of the $j$-th species and $\sigma_{\text{mol}, i, j}$ its cross-section in the $i$-th atmospheric cell. Once the optical depth is computed along all transit chords, the effective area $A_{\text{eff}}(\lambda)$ of the planet can be found from (\citealt{Wardenier2021})

\begin{equation} \label{eq:def_transit_area}
    A_{\text{eff}}(\lambda) = A_{\text{0}} + A_{\text{annu}} \big\langle 1 - e^{-\tau} \big\rangle \big\rvert_{\lambda},
\end{equation}

\noindent with $A_{\text{0}}$ the projected area of the planetary interior and $A_{\text{annu}}$ the area of the atmospheric annulus. The angle brackets denote an average over all photon packets with wavelength $\lambda$.

\subsection{Results}
\label{sect:results1}

The bottom panel of Fig. \ref{fig:three_panels} shows the 10$-$90\% and 1$-$99\% opening angles obtained for the model with \mbox{$R_\text{p} = 10$ $R_\text{Earth}$} and \mbox{$H = 450$} km (solid curves). These were computed from the absorption regions using equation \ref{fig:angle_abs_region}. Because of the wavelength-dependence of the opacities, the opening angle is no longer constant and tends to acquire its highest values inside a line core. This may seem counter-intuitive, because according to equation \ref{eq:opening_angle_final_result}, the opening angle is a monotonically decreasing function of $\sigma_\text{mol}$. However, the solid curves show the opening angle associated with absorption regions that have the \emph{same} values of $\tau_0$ (i.e. \mbox{$\tau_0$ $\sim$ 0.56}), regardless of the wavelength. Consider two wavelengths $\lambda_1$ and $\lambda_2$ with opacities \mbox{$\sigma_1(P_A)$ > $\sigma_2(P_A)$} at a certain pressure $P_A$. If the absorption region of $\lambda_2$ lies around $P_A$, the absorption region of $\lambda_1$ will lie around some pressure \mbox{$P_B < P_A$}, higher up in the atmosphere. To compute the opening angle at $\lambda_1$, we must hence plug $\sigma_1(P_B)$ -- which is potentially smaller than $\sigma_2(P_A)$ -- into equation \ref{eq:opening_angle_final_result}.

The dotted curves in the same panel show the result of evaluating the \citet{Caldas} formula (equation \ref{eq:def_opening_angle}) using the highest pressure ($P_\text{bot}$) and the lowest pressure ($P_\text{top}$) contained by the absorption region at a particular wavelength (see middle panel of Fig. \ref{fig:three_panels}). As illustrated in the middle panel of Fig. \ref{fig:opening_geometry}, this leads to an overestimation of the opening angle, because $P_\text{bot}$ and $P_\text{top}$ do not lie on the same transit chord. Alternatively, if one applies equation \ref{eq:def_opening_angle} to every \emph{individual} transit chord in the absorption region and takes an average (dashed curves), the \citet{Caldas} formula and the radiative-transfer model agree very well. This result demonstrates that $P_\text{top}$ and $P_\text{bot}$ should not be chosen by looking at the \mbox{$\tau_0$ $\sim$ 0.56} level corresponding to the lowest and highest opacity values in the considered bandpass.

As far as the other models are concerned, we note that for all atmospheres with \mbox{$H \leq$ 200 km}, the standard deviation in the opening angle with wavelength is 10 to 15\% of the mean value. For models with \mbox{$H =$ 450 km} and large planetary radii, the relative spread drops to 7\%. Atmospheres with the largest scale height exhibit a very small spread in their opening angle with wavelength, below 2\%.

\section{Opening angles and instrument noise}
\label{sect:noise_floor}

\begin{figure}
\centering
\includegraphics[width=0.5\textwidth]{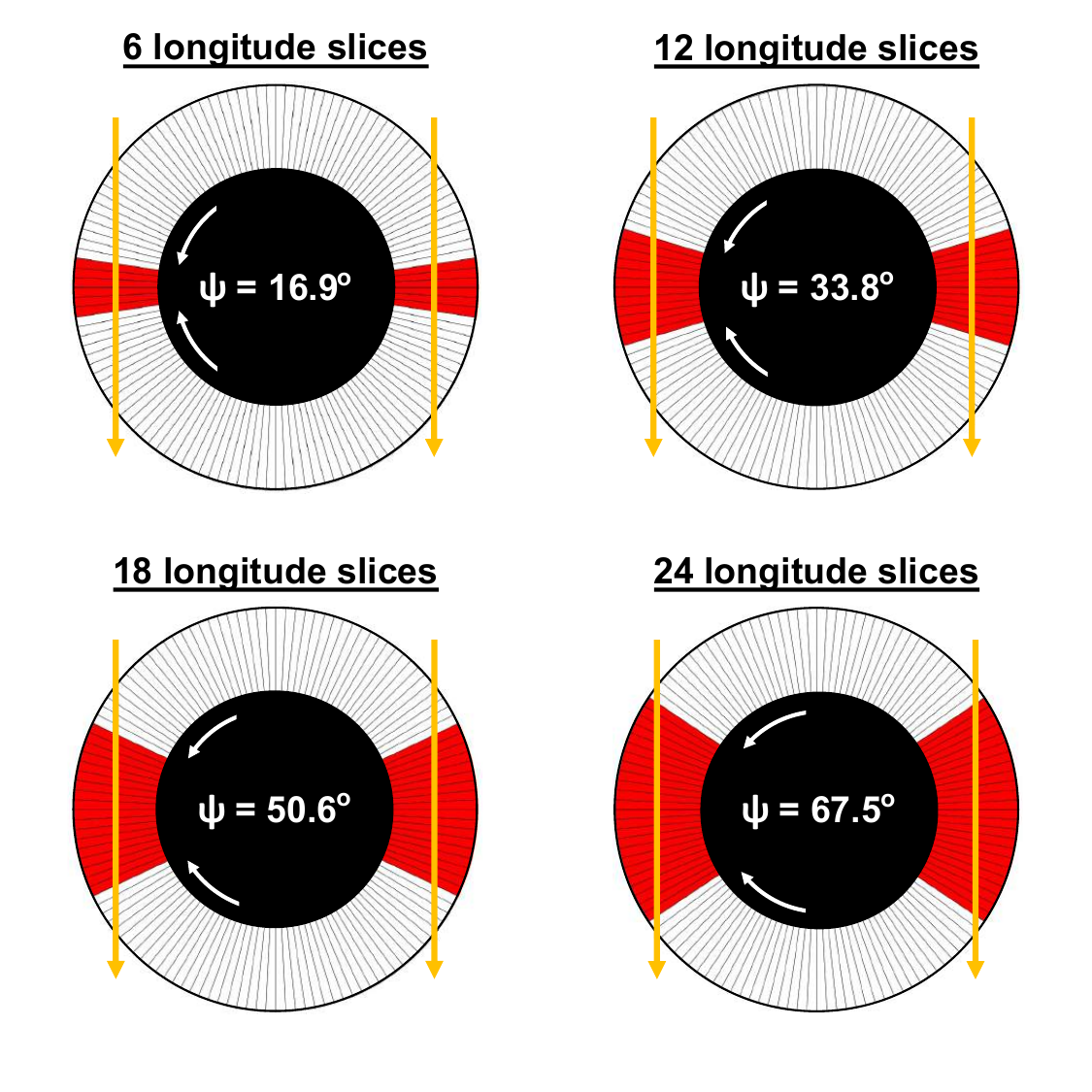}
\vspace{-20pt}
\caption{The equatorial plane of four model atmospheres, where we only preserve the opacities within a certain angle from the terminator (red area). The opacities of the other longitude slices are set to zero (white area). The yellow arrows represent stellar light rays pointing towards the observer. $\psi$ is the angular width of the atmosphere.}
\label{fig:folding_fan}
\end{figure}

In order to make a final, proper comparison between the \citet{Caldas} formula (equation \ref{eq:def_opening_angle}) and our newly derived formula for the opening angle (equation \ref{eq:opening_angle_final_result}), we need to determine what value of $\beta$ is appropriate when estimating the opening angle of an atmosphere. After all, picking a larger value for $\beta$ results in larger values for the opening angle $\psi(\tau_0,\beta)$. As discussed in the Section \ref{sect:prev_work}, the opening angle should be defined such that the opacities \emph{outside} the corresponding absorption region do not have a ``measurable'' impact on the transmission spectrum of the planet. Strictly speaking, this means that the opening angle is not just dependent on the atmospheric structure, but also on the telescope or instrument through which the planet is observed. In this section, we will assume an instrument noise floor of 10 ppm, roughly comparable to that of JWST (\citealt{Beichman2014, Schlawin2021}).

\subsection{Atmospheric models and radiative transfer}
\label{sect:ppm_method}

For each model from Section \ref{sect:radtrans2}, we now seek to determine the number of longitude slices to which an observation with a 10-ppm noise floor is actually sensitive. Unlike \citet{Lacy2020}, we do not switch off the opacities of \emph{individual} longitude slices (see Section \ref{sect:comparison_work} for further discussion). Instead, for each model, we calculate 31 new spectra, where we only preserve the opacities inside the $2n$ longitude slices that are closest to the terminator (with $n = 1,2,3,...,31$). The opacities in all the other slices are set to zero. Fig. \ref{fig:folding_fan} shows what the equatorial plane of the planet looks like when different numbers of longitude slices\footnote{Each longitude slice has an angular width of 360/128 $\approx$ 2.8 degrees.} are ``activated''. Of course, the spectra of the models should converge to the true spectrum of the planet (i.e. $n = 32$) as $n$ increases. To obtain the final transmission spectra, we assume a stellar radius of one solar radius. Once all spectra are computed, we determine the difference between the wavelength-averaged transit depth of the 31 models and that of the full atmosphere ($n = 32$). We then define the opening angle (in degrees) as the smallest value of $(360/64)n$ for which this difference lies below $10^{-5}$.

\subsection{Results}

\begin{figure}
\centering
\vspace{-25pt}
\includegraphics[width=0.5\textwidth]{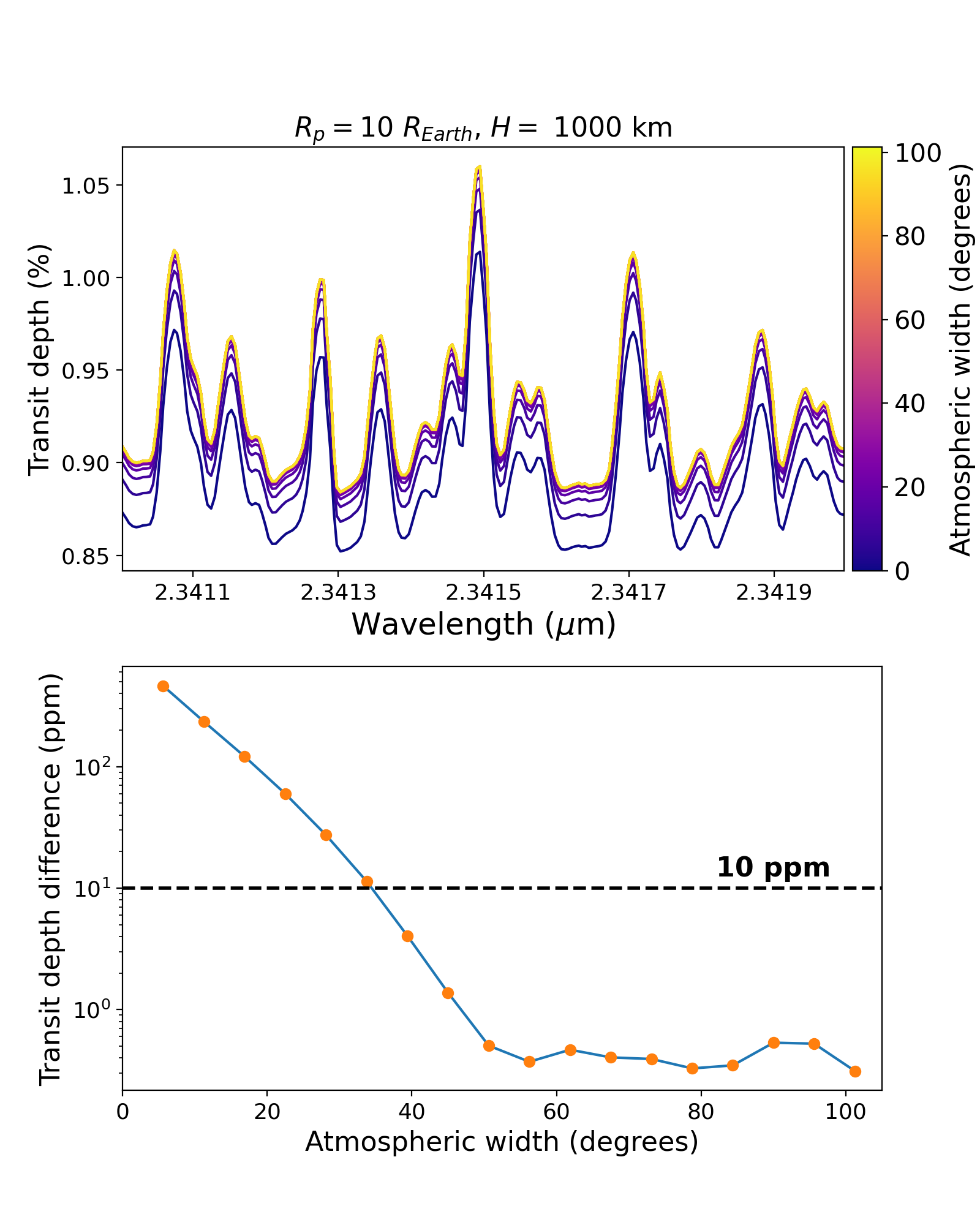}
\vspace{-20pt}
\caption{Radiative-transfer output for the atmosphere with $R_\text{p} = 10$ $R_\text{Earth}$ and $H = 1000$ km. \textbf{Top:} Transmission spectra for models with a different number of activated longitude slices and, therefore, a different atmospheric width. \textbf{Bottom:} The wavelength-averaged difference between the transit depth of the models (orange markers) and the transit depth of the full atmosphere, in which none of the opacities are set to zero.}
\label{fig:ppm_curve}
\end{figure}

The top panel of Fig. \ref{fig:ppm_curve} depicts the transmission spectra of the models with $R_\text{p} = 10$ $R_\text{Earth}$ and $H = 1000$ km. As more longitude slices are activated in the models, the transit depth of the planet increases. At some point, however, the atmospheric width is such that the spectrum can no longer be distinguished from the spectrum of the full atmosphere. The bottom panel of Fig. \ref{fig:ppm_curve} shows the wavelength-averaged difference between the transit depths of the models (orange markers) and the full atmosphere. For this particular combination of planetary radius and scale height, the opening angle is 35 degrees, because models with a larger atmospheric width cannot be distinguished from the full atmosphere above the 10 ppm level. Because of the stochastic nature of the Monte-Carlo code, the simulations also have their own noise floor (below 1 ppm), which is why the curve in the bottom panel does not continue to decrease. 

\begin{figure}
\centering
\vspace{-7pt}
\includegraphics[width=0.5\textwidth]{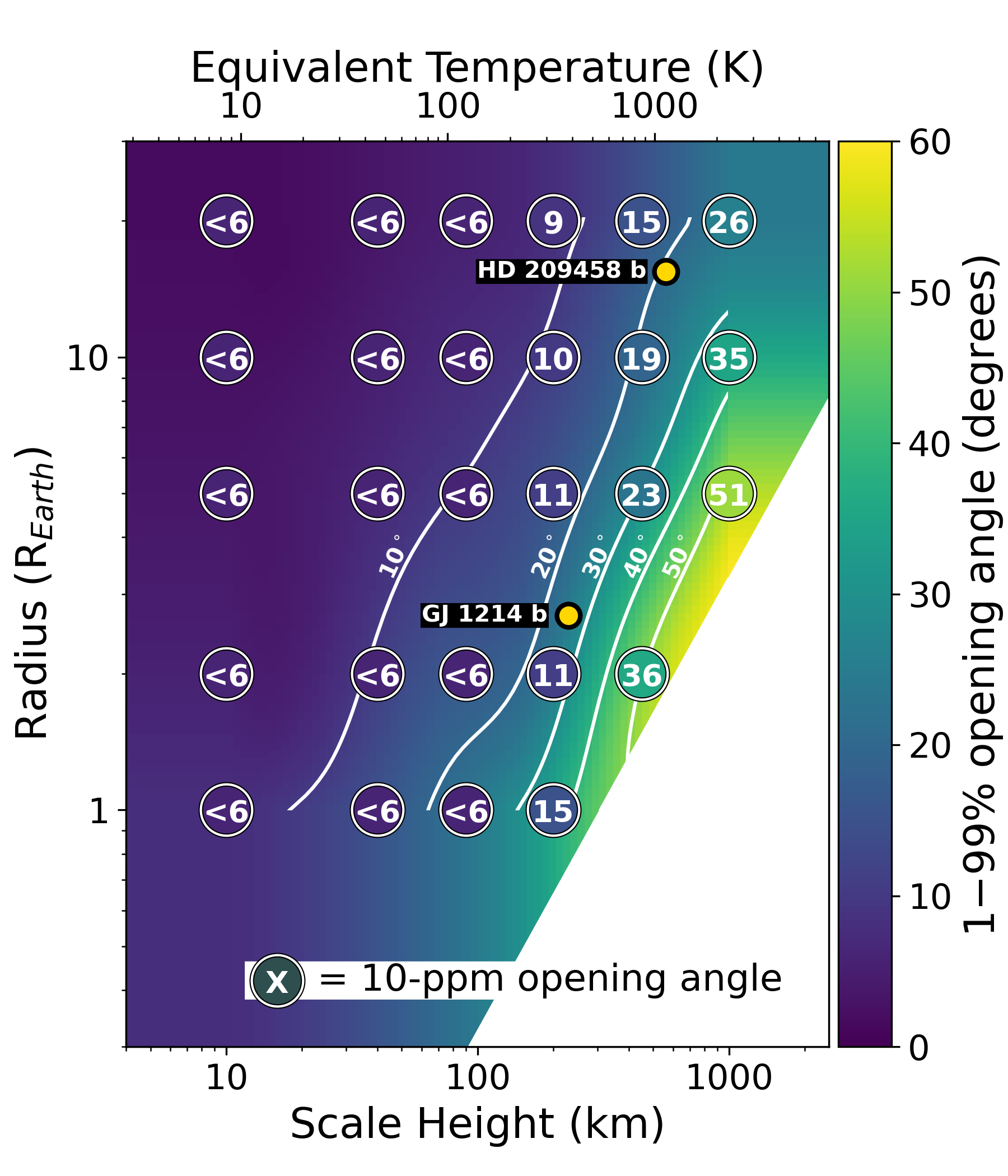}
\vspace{-20pt}
\caption{Opening angles as a function of scale height and planetary radius. The circles contain the opening angles found from the method described in Section \ref{sect:ppm_method}, where we assume a 10-ppm noise floor. The colour map in the background shows the 1$-$99\% opening angles (averaged over all wavelengths from Section \ref{sect:wavelength_dependence}) and was obtained from interpolation. Contour lines are shown in white.}
\label{fig:compare_ppm_to_geo}
\end{figure}

The circles in Fig. \ref{fig:compare_ppm_to_geo} show the 10-ppm opening angles obtained for all the 27 models considered in this work. For all atmospheres with a scale height $H \leq 90$ km, the opening angle is smaller than 6 degrees, because the difference between the spectrum of the $n=1$ model and that of the full atmosphere is already smaller than 10 ppm. In other words, given a noise floor of 10 ppm, one cannot distinguish between the full planet atmosphere and a model with an atmospheric width of only 6 degrees. Therefore, if we want to make a comparison between the opening angles found from the absorption regions (Section \ref{sect:wavelength_dependence}) and those based on the 10-ppm noise floor (this section), we need to focus on the atmospheres with the largest scale heights and planetary radii -- see the top right corner of Fig. \ref{fig:compare_ppm_to_geo}.

For $R$ $\geq$ 5 $R_\text{Earth}$ and $H$ $\geq$ 200 km, the opening angles corresponding to the 1$-$99\% absorption regions are in relatively close agreement with the opening angles obtained from the 10-ppm noise floor. On the other hand, the 10--90\% opening angles (not shown in the figure) are too small to match these values. Based on the models presented here, this suggests that $\beta = 0.01$ is an appropriate value to plug into equation \ref{eq:opening_angle_final_result} when making an estimate of an atmosphere's opening angle $-$ at least when \mbox{$R$ $\geq$ 5 $R_\text{Earth}$} and \mbox{$H$ $\geq$ 200 km}. This is also the regime where most planets with currently known radii and scale heights reside (\citealt{Caldas}). For smaller planetary radii, the 10-ppm opening angles are smaller than those derived from the 1$-$99\% absorption region. Hence, the latter can be seen as an upper limit. 

\section{Discussion}
\label{sect:discussion}

\subsection{Comparison to previous works}
\label{sect:comparison_work}

\begin{figure}
\centering
\vspace{-30pt}
\includegraphics[width=0.5\textwidth]{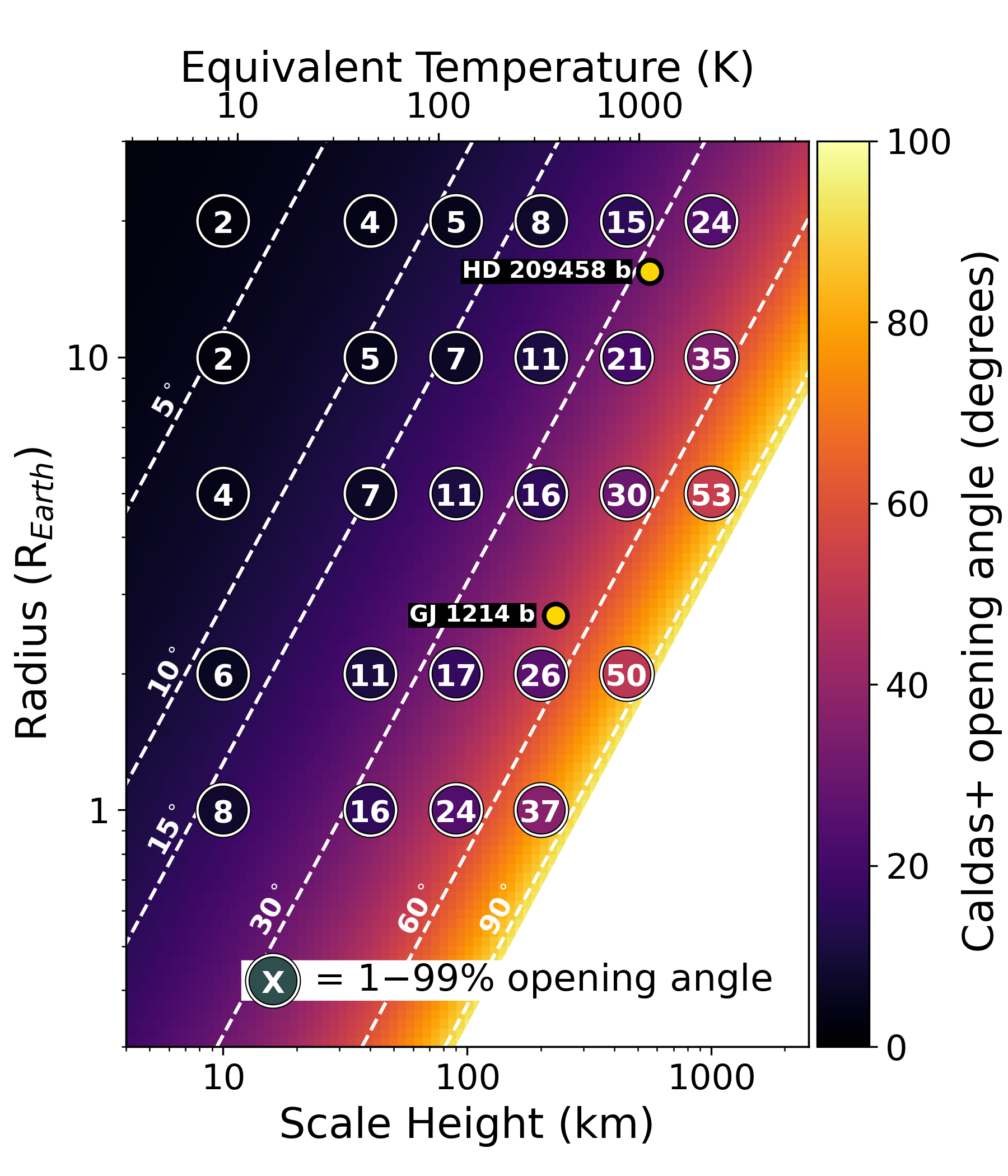}
\vspace{-20pt}
\caption{Opening angles as a function of scale height and planetary radius. The circles contain the 1--99\% opening angles found for all 27 models considered in this work (Section \ref{sect:wavelength_dependence}). The colour map in the background shows the  opening  angles that follow from the \citet{Caldas} formula, assuming $(P_\text{bot}, P_\text{top}) = (10^{-2}, 10^{-5})$ bar. Contour lines are shown in white dashes.}
\label{fig:compare_to_caldas}
\end{figure}

In this work, we derived a formula for the opening angle as a function of opacity and optical depth, instead of the bottom pressure $P_\text{bot}$ and top pressure $P_\text{top}$ probed by a transit observation. This approach is advantageous, because it can be hard to pick values for $P_\text{bot}$ and $P_\text{top}$ such that they lie on the same transit chord. As shown in Fig. \ref{fig:three_panels}, the \citet{Caldas} formula (equation \ref{eq:def_opening_angle}) is correct when applied to a single transit chord. However, when the maximum and minimum pressure probed across a bandpass are plugged into the formula, the opening angle can be strongly overestimated. This is demonstrated in Fig. \ref{fig:compare_to_caldas}. The discrepancy is biggest for planets with small radii and large scale heights.  


In contrast to our study, \citet{Lacy2020} did find a good agreement between their computed opening angles and the \citet{Caldas} formula (see their Fig. 4). Using a 3D radiative-transfer model, the authors applied the same technique as \citet{Molliere2019a} to compute the TCF as a function of longitude. That is, instead of perturbing the opacities across different altitudes, they set the opacities of individual longitude slices\footnote{Technical note (private communication): \citet{Lacy2020} rotated their 3D grid such that its poles coincided with the substellar and antistellar point of the planet. In this way, the latitude slices effectively take on the role of longitude slices, with the benefit that they are equally thick across the entire atmospheric annulus (i.e. there is no singularity at the geographical poles).} equal to zero, and evaluated the difference between the resulting spectrum and the spectrum of the full atmosphere. However, from their work, it is not clear how many longitude slices the authors used, what threshold\footnote{\citet{Lacy2020} refer to a ``non-zero change in the transit spectrum'', which suggests an arbitrarily low noise floor.} (noise floor) they assumed, and how this impacted their results. After all, by lowering the noise floor (i.e. increasing the sensitivity of the instrument), one can always obtain larger values for the opening angle (see Fig. \ref{fig:ppm_curve}).

\subsection{Implications of this study}

The fact that we find smaller opening angles compared to \citet{Caldas} and \citet{Lacy2020} does not mean that 3D modelling becomes less important in the context of retrievals. We find opening angles in the order of 20--25 degrees for (ultra-)hot Jupiters, and this is still a big number compared to the angle over which thermal and chemical gradients can occur. \citet{Caldas}, \citet{Pluriel2020,Pluriel2021} and \citet{Wardenier2021}, for example, present models in which the transition from dayside to nightside chemistry spans less than 10 degrees. Hence, these variations should still be taken into account when interpreting observational data. For cooler planets with smaller opening angles and gradients that are less steep, 1D modelling may be adequate. 

In general, when dealing with variations along the line of sight, it will not be necessary for retrieval models to account for the full 360-degree structure of the atmosphere. Instead, it suffices to have a model that is as wide as the opening angle in the dimension orthogonal to the limb plane. Any opacities outside the opening angle have a negligible effect on the transmission spectrum of the planet. 

\begin{figure}
\centering
\vspace{-20pt}
\includegraphics[width=0.5\textwidth]{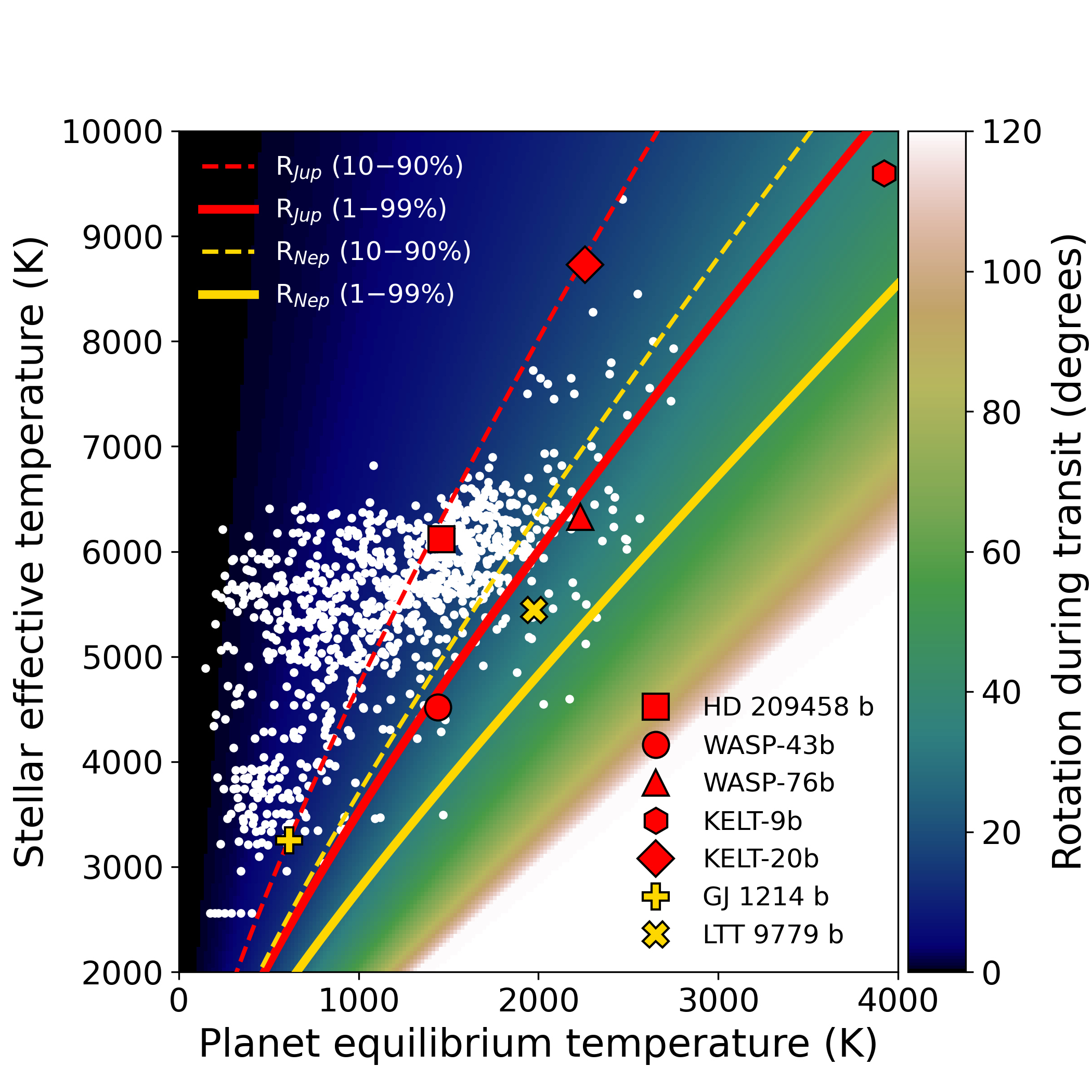}
\vspace{-20pt}
\caption{The angle over which a tidally locked planet rotates during its transit, plotted as a function of its equilibrium temperature $T_\text{eq}$ (assuming zero albedo and full heat redistribution) and the effective temperature $T_\text{eff}$ of the host star. The white bullets show all systems for which the stellar radius and the semi-major axis of the planet orbit are currently known (data acquired from \texttt{https://www.astro.keele.ac.uk/jkt/tepcat/}; \citealt{Southworth2011}). The red and yellow curves show for which ($T_\text{eq}$, $T_\text{eff}$) the planet's rotation angle is equal to the opening angle. Here, we assume two different planet radii (one Jupiter radius and one Neptune radius), and two different values for $\beta$ (see equation \ref{eq:opening_angle_final_result}). To the right of the curves, the rotation angle is larger than the opening angle.}
\label{fig:rotation_angles}
\end{figure}

As discussed in the introduction, absorption regions can be interpreted as (spatial) resolution elements. In the context of tidally locked gas giants, a smaller opening angle allows for (more) independent measurements in and around the equatorial plane of the planet. Figure \ref{fig:rotation_angles} shows the rotation angle of a planet as a function of its equilibrium temperature $T_\text{eq}$ and the effective temperature $T_\text{eff}$ of the host star, assuming it is tidally locked (see Appendix \ref{ap:B} for the relevant equations). The rotation angle is the angle over which the planet rotates during its transit. Given a certain planet radius, there exists a curve connecting all $(T_\text{eq}, T_\text{eff})$ points for which the rotation angle of the planet is equal to its opening angle. To the right of this curve, the rotation angle is \emph{larger} than the opening angle. As shown in Fig. \ref{fig:rotation_angles}, the curve shifts to higher $T_\text{eq}$ values when the planet radius is smaller (red vs yellow). The selected value of $\beta$ impacts the curve in a similar way (solid vs dashed). According to Fig. \ref{fig:rotation_angles}, however, ultra-hot Jupiters such as WASP-76b, WASP-121b and KELT-9b all have rotation angles that are larger than their (1--99\%) opening angles. In the \emph{equatorial plane}, this means that the transit observation probes a completely different atmospheric region at the start of the transit than at the end of the transit. Consequently, measurements of the equatorial jet speed (\citealt{Showman2011,Louden2015}) at the start and the end of the transit can be expected to be fully independent. Such observations would allow for further insights into the atmospheric circulation of (ultra-)hot Jupiters. 

\newpage

\section{Conclusion}
\label{sect:conclusion}

We summarize our most important findings below:

\begin{enumerate}
   \setlength\itemsep{1em}
   \item[$\bullet$] We derived and validated a new formula for the opening angle (equation \ref{eq:opening_angle_final_result}) in the limit of a uniform atmosphere with constant gravity. It does not depend on $P_\text{bot}$ and $P_\text{top}$, the maximum and minimum pressure probed by the transmission spectrum. Instead, the formula depends on the product \mbox{$\sigma_\text{mol}\chi/\tau_0$}, with $\sigma_\text{mol}$ the constant cross-section of the considered species and $\chi$ its number fraction. $\tau_0$ is the total optical depth of a transit chord that crosses the absorption region (typically, $\tau_0 = 0.56$).
   
   \item[$\bullet$] The opening angles of the absorption regions computed with our Monte-Carlo radiative transfer code (\textsc{hires-mcrt}) agree with the new formula (equation \ref{eq:opening_angle_final_result}) in the limit of a grey, uniform atmosphere with constant gravity.
   
   \item[$\bullet$] When we account for the pressure, temperature and wavelength dependence of the cross-sections, we find that the opening angle acquires its maximum value inside a line core.
   
   \item[$\bullet$] The \citet{Caldas} formula (equation \ref{eq:def_opening_angle}) results in an overestimation of the opening angle when $P_\text{bot}$ and $P_\text{top}$ of an absorption region are used. However, when the formula is applied to individual transit chords, it matches the angular extent of the absorption regions very closely.
   
   \item[$\bullet$] The $\beta$ parameter in equation \ref{eq:opening_angle_final_result} governs what segment of the transit chord is subtended by the opening angle. Based on the models presented in this work, we find that $\beta = 0.01$ is a good default choice. The resulting opening angle is the 1--99\% opening angle, which corresponds to segment along which the optical depth increases from 1\% to 99\% of $\tau_0$.
   
   \item[$\bullet$] When we take instrument noise into account (assuming a noise floor of 10 ppm), we find that the opening angle is only bigger than a few degrees for hot planets, with a scale height larger than 100 km. 
   
   \item[$\bullet$] For ultra-hot Jupiters, the typical angle over which the planet rotates during the transit is larger than the opening angle. This implies that measurements of the equatorial jet speed at the start and the end of the transit are fully independent. 
\end{enumerate}

\section*{Acknowledgements}

We thank Brianna Lacy for insightful email correspondences, and Michael Line for interesting discussions. Also, we are grateful to Raymond Pierrehumbert for sharing computing resources. JPW sincerely acknowledges support from the Wolfson Harrison UK Research Council Physics Scholarship and the Science and Technology Facilities Council (STFC). Finally, we thank the anonymous referee for thoughtful comments that helped improve the quality of the manuscript.

\section*{Data Availability}

The data and models underlying this article will be shared on reasonable request to the corresponding author.



\bibliographystyle{mnras}
\bibliography{citations} 




\appendix

\section{The opening angle as a function of the cross-section}
\label{ap:A}

\begin{figure}
\centering
\vspace{-7pt}
\includegraphics[width=0.5\textwidth]{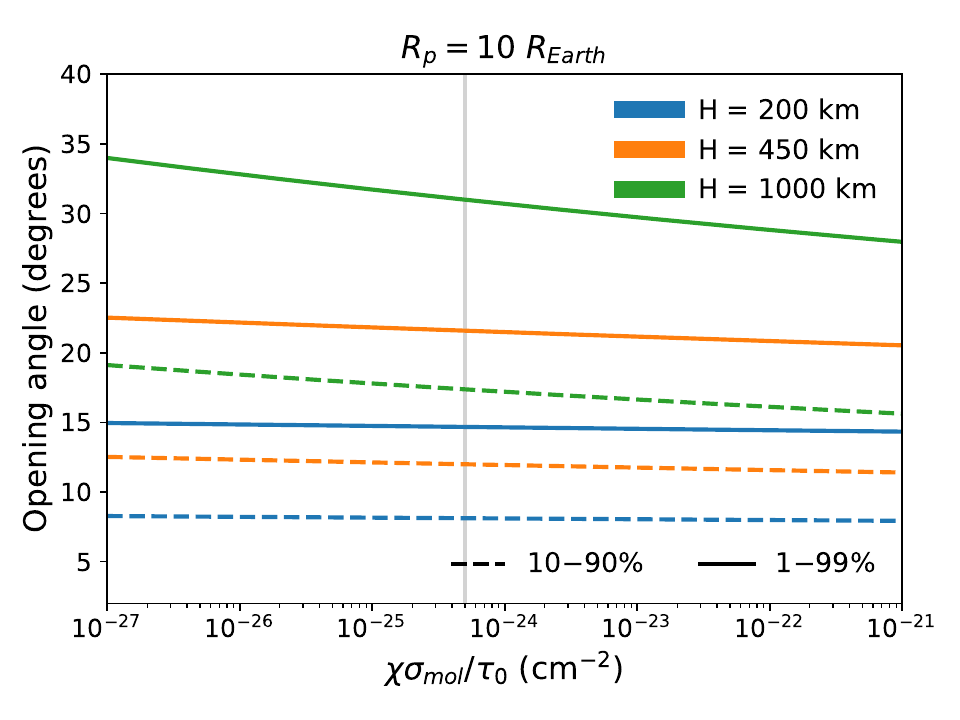}
\vspace{-17pt}
\caption{Plot of the analytical opening angle (equation \ref{eq:opening_angle_final_result}) as a function of $\chi \sigma_\text{mol}/\tau_0$, for planets with a radius of 10 $R_\text{Earth}$. The figure shows the angles for three different scale heights and two values of $\beta$. Here, $\beta = 0.1$ corresponds to the 10--90\% absorption region, while $\beta = 0.01$ gives rise to the 1--99\% absorption region. The grey vertical line denotes the value of $\chi \sigma_\text{mol}/\tau_0$ that is used in this work. Also, we assume a constant gravity $g = 10$ m/s$^2$ and mean molecular weight \mbox{$\mu$ = 2.3 $m_\text{H}$}.}
\label{fig:appendix_plot}
\end{figure}

Figure \ref{fig:appendix_plot} illustrates the dependence of the analytical opening angle from equation \ref{eq:opening_angle_final_result} on the assumed (constant) cross-section and abundance of the absorbing species.

\newpage

\section{Equations behind Figure 9}
\label{ap:B}

In Fig. \ref{fig:rotation_angles}, we plotted the planet rotation angle (assuming tidal locking) as a function of its equilibrium temperature $T_\text{eq}$ and the effective temperature $T_\text{eff}$ of the host star. In this appendix we briefly present the equations behind the colour map and the curves that indicate where the rotation angle is equal to the opening angle. 

For a tidally locked planet, the rotation angle $\psi_\text{rot}$ is given by

\begin{equation} \label{eq:rot_angle1}
    \psi_\text{rot} = 2 \arcsin \big( R_\star/a \big),
\end{equation}

\noindent with $R_\star$ the radius of the host star and $a$ the semi-major axis. Because equation \ref{eq:rot_angle1} does not contain the planet radius, it gives the rotation angle between the middle of the ingress and the middle of the egress. 

Assuming zero albedo and full heat redistribution, the equilibrium temperature of the planet is 

\begin{equation}
    T_\text{eq} = T_\text{eff} \sqrt{ R_\star/2a }.
\end{equation}

\noindent Solving for $R_\star/a$ and plugging the result into Equation \ref{eq:rot_angle1} yields

\begin{equation}
    \psi_\text{rot} = 2 \arcsin \Big( 2 \big(T_\text{eq}/T_\text{eff} \big)^2 \Big).
\end{equation}

\noindent Also, this means that the effective temperature of the host star can be written as

\begin{equation} \label{eq:teff}
    T_\text{eff}(\psi_\text{rot}) = T_\text{eq} \Big/ \sqrt{ \sin\big(\psi_\text{rot}/2\big)/2}
\end{equation}

To compute where the rotation angle is equal to the opening angle, we take three steps. First, we choose a planet radius $R_\text{p}$. Next, we use equation \ref{eq:opening_angle_final_result} to compute the opening angle $\psi_\text{open}$ as a a function of $T_\text{eq}$. The opening angle depends on the equilibrium temperature through the scale height $H$. Furthermore, we assume the same parameter values as in Section \ref{sect:num_ver_formula}. That is, $g = 10$ m/s$^2$, \mbox{$\mu$ = 2.3 $m_\text{H}$} and \mbox{$\chi \sigma_\text{mol}/\tau_0 = 10^{-24.3}$ cm$^{-2}$}. Finally, we use Equation \ref{eq:teff} to compute \mbox{$T_\text{eff}(\psi_\text{rot} = \psi_\text{open})$}.


\bsp	
\label{lastpage}
\end{document}